\documentclass[pre,nopacs,twocolumn,preprintnumbers,notitlepage,amsmath,amssymb,superscriptaddress]{revtex4-2}
\usepackage{amssymb}
\usepackage{amsmath}

\usepackage{color}
\usepackage[pdftex]{graphicx}

\usepackage{mathtools}

\usepackage{makecell}

\usepackage[dvipsnames]{xcolor}

\definecolor{darkblue}{rgb}{0,0,0.6}
\definecolor{darkred}{rgb}{0.6,0,0}
\usepackage[colorlinks=true,urlcolor=darkblue,citecolor=darkblue,linkcolor=darkred]{hyperref}

\usepackage{enumerate}
\usepackage{mathtools}
\usepackage{stmaryrd}

\usepackage{enumitem}

\usepackage{textcomp}
\usepackage{gensymb}

\newcommand{\moy}[1]{\left\langle #1 \right\rangle}

\newcommand{\ex}[1]{\mathrm{e}^{#1}}

\newcommand{\dd}[0]{\mathrm{d}}

\newcommand{\rr}[0]{\boldsymbol{r}}

\newcommand{\kB}[0]{k_{\mathrm{B}}}

\newcommand{\nn}[0]{{\boldsymbol{n}}}

\begin{document}

\title{`Fuelled' motion: phoretic motility and collective behaviour of active colloids}

\author{Pierre Illien}
\affiliation{Rudolf Peierls Centre for Theoretical Physics, University of Oxford, Oxford OX1 3NP, UK.}
\affiliation{Department of Chemistry, The Pennsylvania State University, University Park, PA 16802, USA.}

\author{Ramin Golestanian}
\affiliation{Rudolf Peierls Centre for Theoretical Physics, University of Oxford, Oxford OX1 3NP, UK.}

\author{Ayusman Sen}
\affiliation{Department of Chemistry, The Pennsylvania State University, University Park, PA 16802, USA.}


\begin{abstract}



Designing microscopic and nanoscopic self-propelled particles and characterising their motion has become a major scientific challenge over the past decades. To this purpose, phoretic effects, namely propulsion mechanisms relying on local field gradients,  have been the focus of many theoretical and experimental studies. In this review, we adopt a tutorial approach to present the basic physical mechanisms at stake in phoretic motion, and describe the different experimental works that lead to the fabrication of active particles based on this principle. We also present the collective effects observed in assemblies of interacting active colloids, and the theoretical tools that have been used to describe phoretic and hydrodynamic interactions.

\end{abstract}

\maketitle

%
%
%
%
%
%
%
%
%
%
%

\section{Introduction}

The design of self-propelled particles at the microscopic and nanoscopic length scales is one of the most challenging problems studied in the contemporary physical and chemical sciences. From a fundamental perspective, self-propelled particles constitute examples of nonequilibrium machines, that are able to obtain energy from their environment and to convert it into a mechanical work sufficiently large to overcome the thermal fluctuations and the viscous effects that usually prevail at small length scales. An accurate description of the individual and collective behaviours of synthetic active particles could give a new insight in the understanding of living matter and its self-organisation, which involves out-of-equilibrium objects evolving in complex and fluctuating environments~\cite{Bechinger2016}. From a practical point of view, a better experimental control of the properties of self-propelled particles could lead to the design of smart materials, able to reproduce some functions inspired from cellular biology, such as cargo transport or  chemical sensing~\cite{Wong2016}.

To this purpose, phoretic motion, that has been predicted theoretically in different contexts a long time ago, has received a growing attention. This type of propulsion relies on the interaction between the active particle and a field, like a chemical concentration, an electrostatic potential or a temperature. Local inhomogeneities of this field induce interfacial effects at the surface of the particle, which is driven out-of-equilibrium and set in motion. Using externally imposed field gradients in order to manipulate colloidal particles has been a long-standing problem in transport theory~\cite{Anderson1989}. The case where built-in asymmetries of the particle allow it to generate field gradients, usually named self-phoresis, was investigated more recently, and appeared to be a powerful tool to create self-propelled swimmers. Over the past years, significant technical advances in the fabrication of microscopic particles with complex surface patterns has allowed the emergence of a new field of research, aimed at designing and characterising active particles.

These experiments also allowed the study of collective behaviours that appear in assemblies of interacting active colloids. They originate from the interplay between the nonequilibrium nature of the individual dynamics of the colloids and the long-range interactions that exist between them,  that are mediated by the inhomogeneous field responsible for phoretic motion or by the hydrodynamics effects. Among the different macroscopic effects that were observed, the emergence of cluster-like phases in systems that do not include any built-in attractive interactions is the most striking. Because of the continuous energy conversion at the particle level and of  the complexity of  phoretic and hydrodynamic interactions, the usual tools from statistical physics cannot be used. Deriving a comprehensive theoretical framework that would describe these intrinsically out-of-equilibrium phenomena is a major focus of modern statistical physics.

In this review, following an approach meant to be as  tutorial and non-technical as possible, we present the recent progress in the understanding of phoretic motion and in the collective effects that arise in assemblies of interacting active colloids, both from a theoretical and an experimental perspective. We start from the basic physical considerations that identify the different features leading to self-propulsion of microscopic particles in fluid environments. We list the main phoretic effects that have been studied, and we show how the velocity of an active colloidal particle can be estimated mathematically in the particular case of self-diffusiophoresis. We then show how these theoretical predictions have been used as guidelines to design and make phoretic particles and how their motion was characterised experimentally, both for particles self-generating the field gradients around them or driven by some external manipulation. The experimental section ends with a review of the collective effects that are usually observed in assemblies of active particles. We then highlight the theoretical difficulties that arise in the study of interacting phoretic particles, and present the recent breakthroughs that permitted to study analytically and numerically the effects of phoretic and hydrodynamic interactions in the emergence of large-scale collective effects.

\section{The basics of phoretic motion}

\subsection{Ingredients for propulsion and basic examples}

Relying on a few basic physical considerations, one can identify the ingredients that are necessary to achieve propulsion of a microswimmer in a fluid. First, the particle needs to convert energy into propulsion, which is the signature of the nonequilibrium nature of the process. This energy can be provided to the motile particle by an external manipulation of the system, or it can be directly pumped into the surroundings of the particle. In other words, the particle needs to interact with a field, that we will generally denote by $\phi$, and whose nature will be specified later on. Secondly, in order to make the displacement directed (at least partially) the symmetry of the system needs to be broken in some way. The field with which the particle interacts needs to be inhomogeneous, which can be achieved either by imposing a gradient over the system or  by making the motile object asymmetric.

\begin{figure}
\begin{center}
\includegraphics[width=\columnwidth]{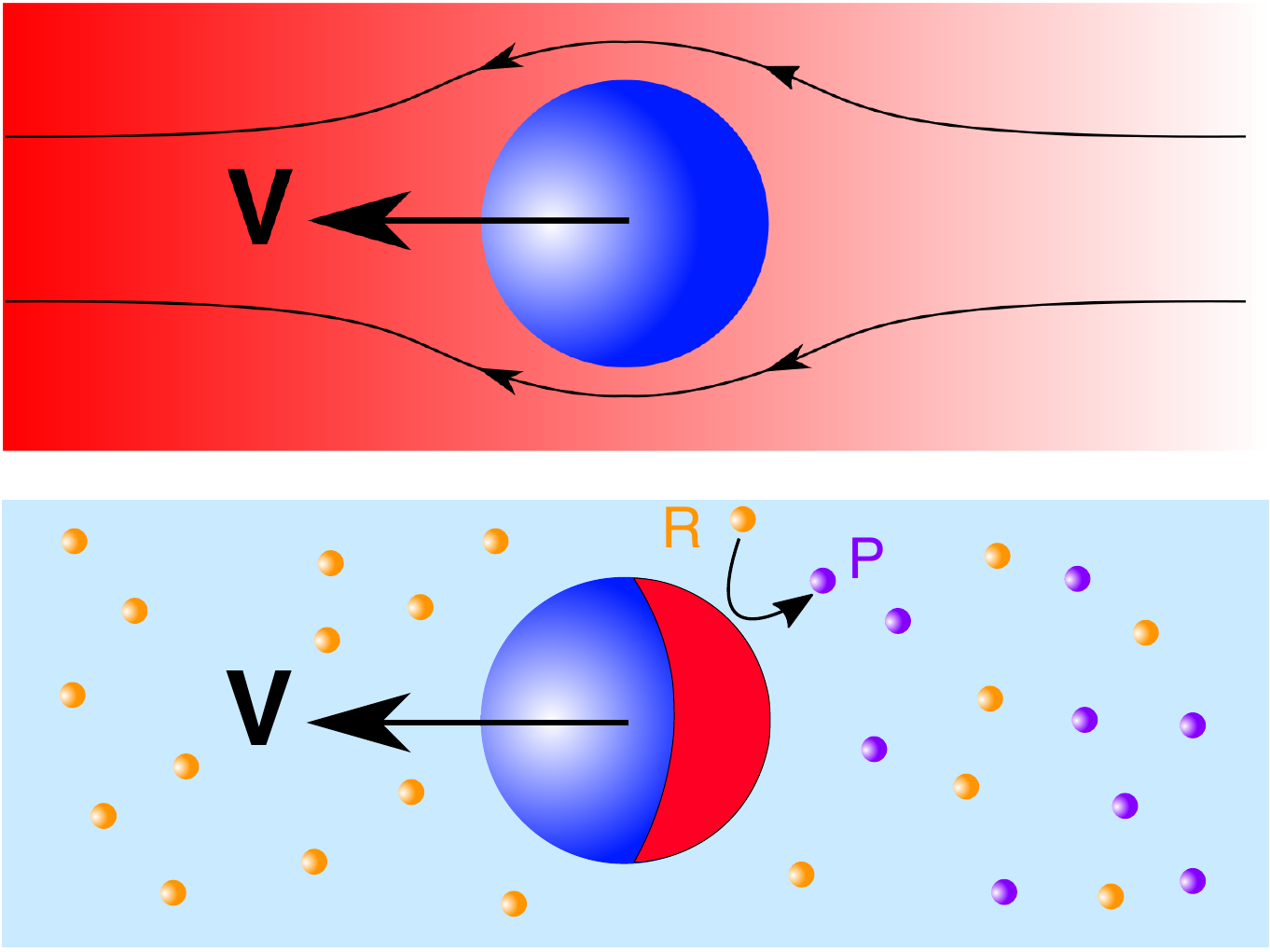}
\caption{Top: A symmetric colloid is placed in an externally imposed temperature gradient, thermophoresis drives the colloid along the gradient. Bottom: A self-diffusiophoretic colloid with a built-in asymmetry: half of its surface is coated with a catalyst that transforms the surrounding reactant molecules (R) into a product (P).}
\label{simple_examples}
\end{center}
\end{figure}

In order to be more specific, let us give two different examples of phoretic motion. First consider a spherical droplet of liquid placed in another liquid (Fig. \ref{simple_examples}). If the system is subject to a temperature gradient, the surface tension, which is usually a strongly temperature-dependent quantity, will be inhomogeneous at the interface. This will result in stresses and flows near the surface of the droplet, and to a net displacement of the droplet along the direction of the gradient~\cite{Anderson1989}. In this case, the temperature plays the role of the field $\phi$, and the symmetry breaking is externally imposed by the experimental conditions.

Let us then consider a spherical solid particle, placed in a reactant solution, and which is half-coated with a solid catalyst that transforms the reactant into a product (Fig. \ref{simple_examples}). Because of the catalytic reaction that takes place at the surface of the sphere, there will be more products molecule on the coated side of the molecule. The interactions between the swimmer and the solute particles become asymmetric and cause a net propulsion of the motile particle~\cite{Golestanian2005}. Here, the field with which the swimmer interacts is the concentration of solute molecules. As opposed to the first example, the symmetry breaking of this field is caused by the particle itself, and not by a macroscopic constraint. This example falls in the class of \emph{self-phoresis}, that we will describe in details in this review.

Based on these general concepts, propulsion induced by phoretic effects has been predicted theoretically and demonstrated experimentally using different field gradients and different kinds of particles:
\begin{enumerate}
  \item[(i)] \emph{electrophoresis}: the particle responds to an electric field gradient.
  \item[(ii)] \emph{thermophoresis}: the particle responds to an inhomogeneous temperature field.
  \item[(iii)] \emph{osmophoresis}: a porous object (such as a lipid vesicle~\cite{Nardi1999a}) is placed in a solute concentration gradient and driven by asymmetric osmotic pumping.
  \item[(iv)] \emph{diffusiophoresis}: the colloid interacts with smaller solute molecules placed in the bulk, that are inhomogeneously distributed.
  \item[(v)] \emph{acoustophoresis}: the particle is placed in an ultrasound wave and is driven by a gradient in the acoustic pressure.
\end{enumerate}
For all these mechanisms, the field gradients can in principle be externally imposed or self-generated by the motile particle.

\subsection{Estimating the colloid velocity}

Going further into the details of the modelling, we now quantify the displacement of the colloidal particle, and, more precisely,  relate its velocity to the spatial modulations of the  field $\phi$. Generically, the motion of the colloid is due to the imbalance of osmotic effects within the solid/liquid interfacial structure at the surface of the object. In order to make this idea more concrete, we will focus on the generic example of diffusiophoresis, for a neutral solute with an inhomogeneous concentration, independently of the origin of this inhomogeneity.

If there is a gradient of solute concentration along the surface of the colloid (parametrized by a unit vector ${\boldsymbol{n}}$) and in the dilute limit,  the concentration field has the Boltzmann-weight form $c(\boldsymbol{r}) = c_0({\boldsymbol{n}})\exp(-\psi(r)/\kB T) $ at any point $\rr$ outside of the colloidal particle, where $\psi$ is the interaction potential between the solute molecules and the surface of the colloid. In the stationary state, the Navier-Stokes equation (NS) reduces to $-\eta \nabla^2\boldsymbol{v} = -\nabla p -c \nabla \psi$, where $\eta$ is the viscosity of the solvent, $\boldsymbol{v}$ the velocity field, $p$ the pressure field and $-c \nabla \psi$ is the osmotic force originating from the solute concentration inhomogeneities, {and corresponds to the gradient of $c$ in the radial direction}. The projection of the NS equation along the radial direction shows that the quantity $p-\kB T c$ is a constant, which physically represents the balance between the hydrostatic and osmotic pressures. Using this relation in the tangential projection of the equation and solving for the velocity field far from the surface of the colloid yields the following expression for the slip velocity at a given point of the surface of the colloid $\boldsymbol{r}_\text{s}$ {\cite{Anderson1989}},
\begin{equation}
\label{ }
\boldsymbol{v}_\text{s} (\boldsymbol{r}_\text{s}) = - \frac{\kB T}{\eta} \left( \int_0^\infty z(\ex{-\psi(z)/\kB T}-1)\dd z\right) \nabla_{\parallel} c(\boldsymbol{r}_\text{s}),
\end{equation}
where the gradient $\nabla_{\parallel}$ corresponds to the tangential direction at the surface of the colloid. This result was recently extended in different directions. First, we should note that this derivation does not take into account the advection of the solute molecules, which is acceptable as long as the interaction layer is much smaller than the diameter of the particle. In recent analytical studies, Michelin and Lauga considered the effect of advection for different values of the P\'eclet number, which compares the diffusive and advective timescales~\cite{Michelin2014}. In the particular case where the surface of the colloid is engineered to let the solvent have a finite slip length at its surface, it was shown that a very large amplification of $\boldsymbol{v}_\text{s}$ can occur~\cite{Ajdari2006}.

Similar derivations can be carried out for the other phoretic mechanisms, and the slip velocity takes the generic form $\boldsymbol{v}_\text{s} (\boldsymbol{r}_\text{s}) = \mu(\boldsymbol{r}_\text{s}) \nabla_{\parallel} \phi(\boldsymbol{r}_\text{s})$ where $\mu(\boldsymbol{r}_\text{s})$, is the local phoretic \emph{mobility}. This quantity depends on the surface properties of the colloid, that are usually finely tuned in the experimental designs. Finally, the velocity of the centre of mass of the colloid, denoted by $\boldsymbol{v}$, can be deduced from the slip velocity using the reciprocal theorem for low-Reynolds number hydrodynamics~\cite{Golestanian2007}. We emphasise that phoretic motion is only due to a non-vanishing slip velocity along the surface of the colloid, and does not originate from a net external  force or torque. This propulsion mode is then often called \emph{force-free} or \emph{torque-free}.

\subsection{The case of self-diffusiophoresis}

The expressions derived in the previous section show  that the velocity of the colloid is non-zero when there exist field gradients around it, independently of what actually creates them. As it was exemplified above, the colloidal particles can self-generate the field inhomogeneities in their environment, by having an inhomogeneous surface activity. In the simple case where the  catalytic part of the surface acts a source (or a sink) of smaller particles, the concentration field around the colloid obeys the stationary diffusion equation $D\nabla^2 c=0$, where $D$ is the diffusion coefficient of solute molecules, that needs to be solved with the boundary condition $-D \boldsymbol{n} \cdot \nabla c(\boldsymbol{r}_\text{s}) = \alpha(\boldsymbol{r}_\text{s})$. The coefficient $\alpha(\boldsymbol{r}_\text{s})$ is the local \emph{activity} of the surface of the colloid, whose sign and magnitude describe the production or destruction rate of the solute molecules, and which is determined by the built-in surface properties of the object.

The order of magnitude of the overall velocity of the colloid can then be estimated as~\cite{Golestanian2007}
\begin{equation}
\label{ }
V\sim \alpha \mu /D,
\end{equation}
and grows linearly with the activity and mobility coefficients. With a more detailed resolution of the set of equations given above, the velocity of the colloid can be computed for different object shapes (spheres, rods...) with different surface patterns for the mobility and the activity. This theoretical approach then gives a guideline to design and optimise these active particles.

\subsection{The role of thermal fluctuations}

The theory presented above is purely deterministic and does not take into account  the thermal fluctuations, that are expected to play a significant role at the microscopic and nanoscopic length scales. Because of the stochastic nature of the dynamics, it is difficult to control the directionality of the colloid, and the evolution of its position needs to be described using statistical averages. Let us consider an isolated active colloid whose velocity has a fixed magnitude $V$ but an orientation $\boldsymbol{u}$ which is randomised over a time scale $\tau_\text{r} \sim 8\pi\eta R^3/\kB T$ (for a solid sphere of radius $R$), such that the orientation auto-correlation function satisfies $\langle \boldsymbol{u}(t)\cdot \boldsymbol{u}(0) \rangle \sim \ex{-t/\tau_\text{r}}$. It is straightforward to show that the mean-square displacement (MSD) of the colloid obeys~\cite{Howse2007}
\begin{equation}
\label{ }
\langle\boldsymbol{r}^2\rangle = 6 D_0 t +V^2 \tau_\text{r}^2\left[\frac{t}{\tau_\text{r}} + \frac{1}{2}(\ex{-2t/\tau_\text{r}}-1)\right],
\end{equation}
where $D_0=\kB T /(6\pi\eta R)$ is the diffusion coefficient given by the Stokes-Einstein relation. This equation reveals the rich behaviour of the stochastic dynamics of the active colloid: at short times ($t\ll\tau_\text{r}$), the `active' contribution gives a ballistic contribution to the MSD; at long times ($t\gg\tau_\text{r}$), it gives a diffusive contribution, which results in an effective diffusion coefficient that depends explicitly on the propulsion velocity, and that reads $D_\text{eff} = D_0+ \frac{1}{6} V^2 \tau_\text{R}$.

In the particular case of a self-diffusiophoretic propulsion mechanism, a more thorough analysis can be carried out to compare the rotational diffusion timescale $\tau_\text{r}$ with the timescale related to the diffusion of the solute, given by $\tau_\text{d} = R^2/D_0$ (where $D_0$ is the diffusion coefficient of indiviual solute molecules), and the timescale related to the hydrodynamic effects $\tau_\text{h} = R^2/\nu$, where $\nu=\eta/\rho$ is the kinematic viscosity of the solvent. The interplay between the mechanisms controlled by these different timescales gives rise to many different regimes (inertia, propulsion, anomalous diffusion)~\cite{Golestanian2009}. In the long-time limit, the effective diffusion coefficient includes different contributions and was shown to be a non-monotonic function of the particle radius, which gives a prescription to optimise the design of active colloids in experimental realisations.

\section{Designing and making phoretic swimmers}

Because of a number of technological limitations, the study of phoresis and self-phoresis has remained essentially theoretical for many decades. The experimental realisations of self-propelled swimmers was made possible by the recent progress in micro- and nano-fabrication techniques, and in the possibility to build easily and at relatively low costs colloidal particles with complex and asymmetric functionalisations. In this section, we present the typical design of the  phoretic swimmers that were built and characterised over the past decade, as well as the collective effects that were observed in large assemblies of such objects.

\subsection{Self-generated field gradients}

\begin{figure}
\begin{center}
\includegraphics[width=\columnwidth]{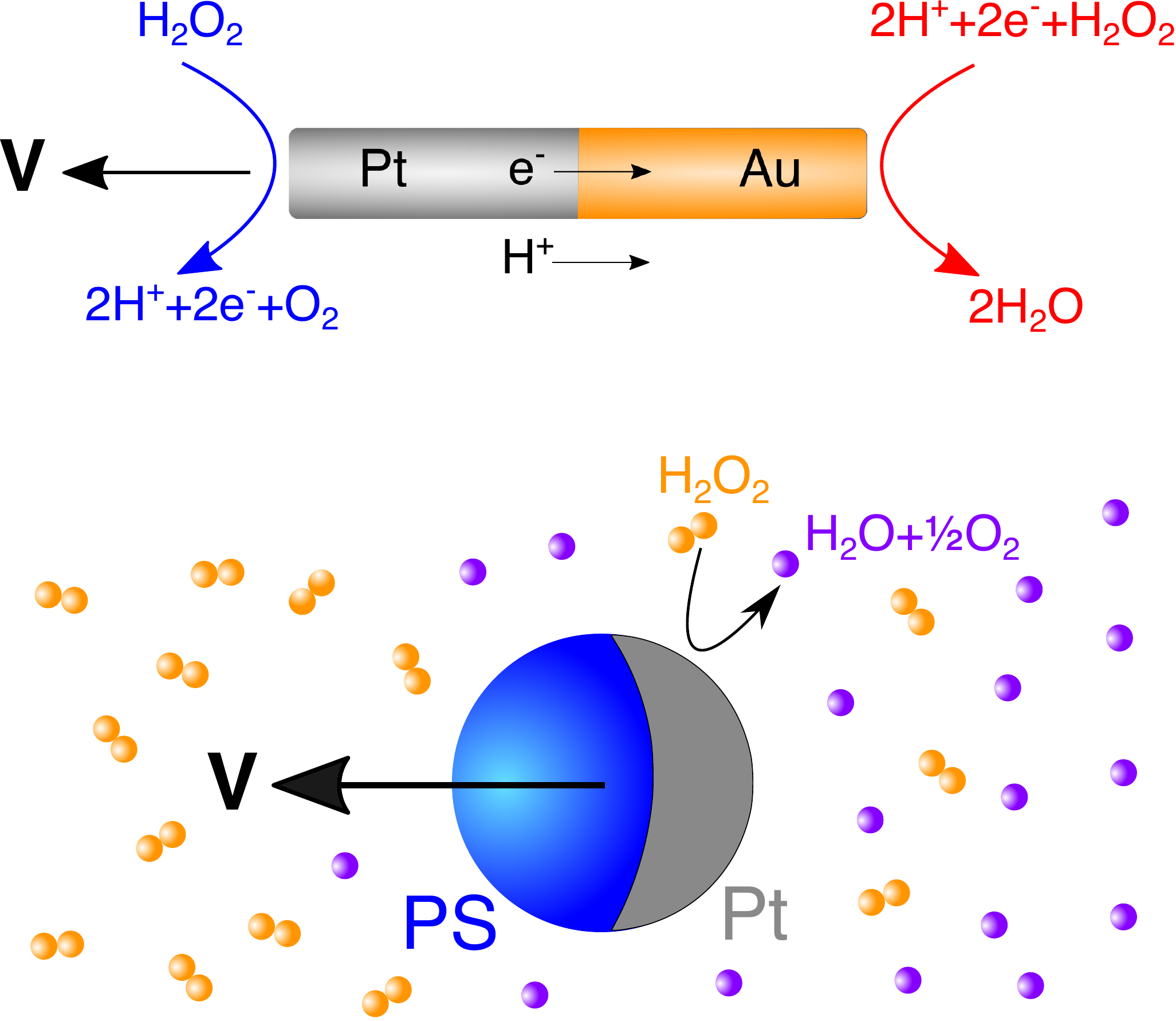}
\caption{Top: Bimetallic rod in a solution of hydrogen peroxide and driven by a self-electrophoretic mechanism~\cite{Paxton2004,Paxton2006}. Bottom: polystyrene-platinum Janus particle driven by a self-diffusiophoretic mechanism: the hydrogen peroxide (represented by the orange dimers) are decomposed into oxygen and water at the platinum side of the colloid~\cite{Howse2007}.}
\label{Paxton_Howse}
\end{center}
\end{figure}

The first example of a synthetic micromotor with a built-in asymmetry and able to move in a self-generated gradient was given by Sen, Mallouk and co-workers in 2004~\cite{Paxton2004,Paxton2006}, who built bimetallic Au-Pt rods (470 nm in diameter and made of two 1 $\mu$m-long metallic cylinders). {These motors were propelled by a self-electrophoretic effect: the oxidation of hydrogen peroxide is catalysed by the Pt end of the rod, and the protons and electrons produced by this reaction are consumed by the reduction of hydrogen peroxide at the Au end of the rod (see Fig. \ref{Paxton_Howse}). This results in an electron current through the rod, and a flow of protons through the solution along the surface of the rod. The latter creates a fluid flow that results in a very fast propulsion, with a velocity that can be greater than 10 body lengths per second.} These cylindrical motors were made using electrochemical deposition in pore templates. The process allows the creation of uniform rods, whose dimensions can be finely controlled. 

Relying on their robust transport properties, it was shown that these bimetallic rods were able to collect and carry small passive particles at one of their ends~\cite{Wang2013}, which opens the way to building autonomous self-propelled rafts for cargos. They were also shown to exhibit directed motion in a gradient of fuel~\cite{Hong2007}, thus providing a first example for chemotaxis in a nonbiological system. This propulsion mechanism was also applied to drive the rotational motion of bimetallic gears~\cite{Catchmark2005}.

Following on theoretical predictions for propulsion due to self-diffusiophoresis~\cite{Golestanian2005}, the first spherical active colloids were synthesised in 2007 by Golestanian and co-workers~\cite{Howse2007}. Micrometer-sized polystyrene spheres were half-coated with platinum and placed in a solution of hydrogen peroxide. The colloid is able to catalyse the decomposition of H$_2$O$_2$ at the platinum side only, which creates a fuel gradient around the particle and results in self-propulsion (Fig. \ref{Paxton_Howse}). A recent investigation, combining experimental and theoretical results, revealed that the details of the propulsion mechanism are quite complex, and that self-electrophoretic effects would also need to be taken into account to give a full and consistent picture~\cite{Ebbens2014}. The typical velocity of the particle is of the order of a few diameters per second. It was recently shown that the presence of boundaries (plane walls, channel-like geometries) could enhance the persistence of the displacement of the particle and therefore allow to steer them along given trajectories {\cite{Das2015,Simmchen2016}}.

This design inspired other realisations, and led to the creation of different types of Janus particles with different types of reactants and coatings, among which we can cite the use of hematite, that acts as a catalyst for the decomposition of hydrogen peroxide only when illuminated with blue light, and therefore provides a versatile system where self-propulsion can be switched on and off~\cite{Palacci2013}.

\begin{figure}
\begin{center}
\includegraphics[width=7cm]{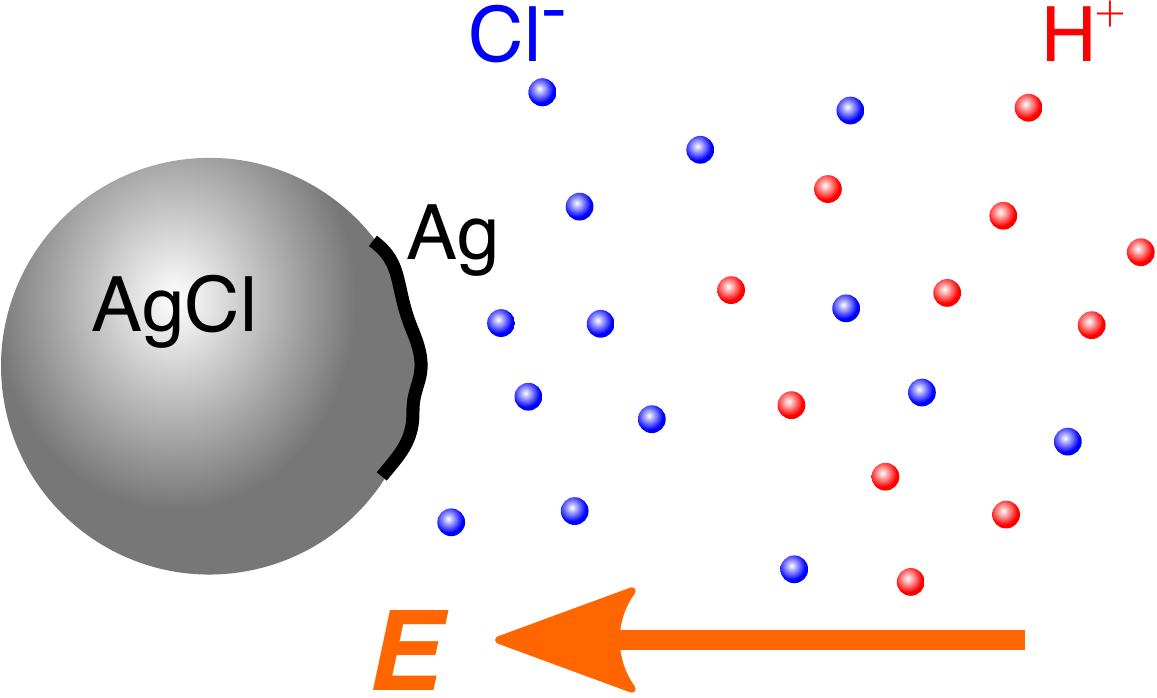}
\caption{Mechanism for electrolyte self-diffusiophoresis. In the presence of UV light, the following reaction takes place at the surface of the particle: AgCl + 2H$_2$O + $h\nu$ $\rightarrow$ 4Ag + 4H$^+$ + 4Cl$^-$ + O$_2$. The difference in the diffusivities of the H$^+$ and Cl$^-$ ions yields an inhomogeneous charge distribution around the particle, and a net electric field~\cite{Ibele2009,Ibele2010c}. In the presence of hydrogen peroxide, the solid layer Ag can be transformed back into silver chloride.}
\label{agcl}
\end{center}
\end{figure}

In the situation where the products generated by the catalytic reactions taking place at the surface of the colloid are charged, an additional electrophoretic effect can propel the particle. This was evidenced by making micrometer-sized AgCl paricles~\cite{Ibele2009,Ibele2010c}. Under UV illumination, silver chloride is reduced to Ag, while H$^+$ and Cl$^-$ ions are released in the solution.  Given the difference in their sizes, the H$^+$ ions will diffuse much faster than the Cl$^-$ ions, thus creating an inward radial electric field around the AgCl particle. Because of the surface inhomogeneities, the particle bears a built-in asymmetry. The resulting imbalance in the distribution of the ions creates a net electric field, in response to which the particle moves (Fig. \ref{agcl}).
The slip velocity of the particle, which results from a gradient of electrolyte concentatrion, therefore has both diffusiophoretic and electrophoretic contributions. Morevover, the addition of hydrogen peroxide in the solution allows the reverse oxidation, which can lead to oscillatory motions of the AgCl particles. It was also shown that titanium dioxide (TiO$_2$) particles~\cite{Hong2010a}  and Ag$_3$PO$_4$ particles~\cite{Duan2013} can move thanks to analogous photoactive mechanisms.

In addition to electrophoresis and diffusiophoresis effects, thermophoretic effects have also been employed to build self-propelled colloids. It was shown that micrometer-sized silica beads half-coated with a gold cap and irradiated with a laser beam could move at relatively high speeds, up to 10 particle diameters per second~\cite{Rings2010}. The gold-coated side of the particle absorbs more light than the silica side, and the colloid plays the role of an asymmetric heat source. The local temperature gradient and the subsequent viscosity inhomogeneities drive the motion of the particle. Despite the efficiency of this propulsion mechanism, the relatively high laser intensities that are required could  be limiting in practical applications. The situation where the colloid evolves in a binary liquid close to a critical point then raised some interest: with smaller light intensities, the  heating at the metallic side of the colloid can cause a local phase separation. The resulting concentration gradient drives the particle, and this diffusiophoretic mechanism actually dominates over the thermophoretic contribution.This phenomenon has been evidenced both experimentally~\cite{Volpe2011,{Buttinoni2013}} and theoretically~\cite{Wurger2015}.

Another class of self-propulsion mechanisms relies on the existence of Marangoni stresses (generated by surface tension gradients) at the surface of the particle. Thutupalli \emph{et al.} studied droplet of waters in oil, with mono-olein, an organic molecule, as a surfactant~\cite{Thutupalli2011}. Bromine molecules contained inside the droplet play the role of the fuel: when a fraction of the surfactant is brominated, the surface tension at the water-oil interface is locally modified, and Marangoni flows are generated. In this situation, the droplets do not bear a built-in asymmetry, and self-propulsion relies on a spontaneous symmetry breaking. Another experimental realisation, supported by theoretical arguments, suggested that the motion of the droplet can be driven by a different mechanism: under large P\'eclet number conditions, the solubilisation of water from the droplet can create gradients of micelles, and the droplet can be driven by a combination of phoretic and Marangoni effects~\cite{Izri2014}. This system is fully biocompatible, and allows the transport of biological cells, opening the way to many practical applications.

\subsection{Externally driven particles}

In contrast with particles that generate themselves the field gradients around them, microswimmers can also be driven by external fields imposed by the experimental conditions. This strategy has been applied to design another type of metallic nanorods that move thanks to an acoustophoretic mechanism~\cite{Wang2012}. It was shown that these nanorods can levitate in the nodal plane of externally-generated ultrasonic standing waves, and reach propulsion velocities up to 200 $\mu$m/s, which correspond to approximately 100 body lengths per second. The shape asymmetry of the rods, that originates from the fabrication technique employed, creates an asymmetric distribution of the acoustic pressure, responsible for self-propulsion. This mechanism is independent of the chemical environment of the rod, and propulsion can therefore be observed in solutions with very high ionic strengths. This property, as well as the fact that acoustic waves could be less destructive than optical or magnetic manipulation,  suggested that these nanorods could be used in biological environments. Their bio-compatibility was successfully demonstrated by making them propel inside living cells~\cite{Wang2014}.

Another propulsion mechanism relying on the interaction with an external field and on a spontaneous symmetry breaking was considered by Bricard \emph{et al.}, who studied dielectric colloids placed in a conducting fluid in the presence of an external electric field. When the magnitude of this field exceeds a threshold, the charge distribution at the surface of the colloid is unstable, which causes a symmetry breaking~\cite{Bricard2013}. Under the effect of the electric field, the colloid can then move at very high speed (up to 200 particle diameters per second).

We should also mention another swimming strategy that does not rely on phoretic effects, and that has been widely used to design microswimmers with internal degrees of freedom. In the limit of low Reynolds number, the inertial effects are negligible, and the equations for momentum conservation have symmetries that imply that the fluid obeys kinematic reversibility. Consequently, any reciprocal sequence of deformation of the swimmer cannot result in a net displacement. This property, usually referred to as the `scallop theorem'~\cite{Purcell1977}, represents a difficulty in the design of periodically actuated swimmers. It can be overcome by designing swimming strategies that include non-reciprocity, coming from the actuation of the internal degrees of freedom or from a sufficiently complex geometry. An object made of three spheres with links whose lengths are actuated in a non-reciprocal fashion then constitutes a minimal model for a low-Reynolds-number swimmer, that can be studied analytically~\cite{Najafi2004}.  In a first attempt to build and characterise experimentally a swimmer of this kind, Dreyfus \emph{et al.} built chains of micrometer-size colloids, linked by DNA strands~\cite{Dreyfus2005}. The non-reciprocal actuation of the `flagellum' using a time-dependent magnetic field allows its propulsion, and it was successfully used to carry a red blood cell to which it is attached. Relying on a similar external magnetic actuation, Tierno \emph{et al.} designed a simpler particle made of two colloids of different sizes~\cite{Tierno2008}. In this system, the swimmer does not have enough internal degrees of freedom to achieve self-propulsion, and the time reversibility breaking is achieved by the presence of a solid wall, which causes an asymmetry in dissipation. Later on, Ghosh and Fisher designed and characterised glass nanostructured propellers which have the shape of a spherical head attached to a corkscrew-like tail, and which could achieve propulsion by responding to a magnetic field~\cite{Ghost2009}.

\subsection{Self-assembly and collective effects}

In addition to their remarkable individual self-propulsion properties, phoretic particles can also display many surprising collective effects. They result from the complex dynamics of the particles at the individual level, combined with the interactions that exist between them, and that can have many different origins: excluded-volume interactions, hydrodynamic interactions mediated by the solvent, or long-ranged  interactions mediated by the field responsible for  phoretic motion (temperature, concentration, electric field...).  In this section, we present the different classes of collective effects that have been observed experimentally, and we will review the different theoretical approaches that could account for them in the last section of this review.

One of the most striking features of assemblies of active colloids is the formation of dynamic  clusters of macroscopic size, separated by gas-like dilute regions. This clustering phenomenon is surprising as it appears in the absence of any built-in attractive interactions, and is a consequence of the out-of-equilibrium self-propulsion dynamics. It can be understood intuitively as follows: because of the persistence of the colloid trajectories and of their exclusion interactions, several particles can be stopped when their propulsion direction face one another. Large-scale structures may then appear when other particles join elementary clusters, which is possible when the typical rotational diffusion time (namely the time after which each colloid looses memory of its orientation) is very large compared to the time between inter-particle collisions, controlled by the typical propulsion velocity and the density.

This clustering was observed with self-diffusiophoretic platinum-gold particles in H$_2$O$_2$ \cite{Theurkauff2012},  with SiO$_2$-carbon particles in a near-critical binary mixture~\cite{Buttinoni2013}, and  with colloids carrying small hematite cubes~\cite{Palacci2013}. It was observed that the clusters are dynamical: they can expel particles or recruit new ones, rotate or merge to form larger clusters. {The dominant physical mechanisms at stake in the cluster formation depend on the details of the experimental systems: for Janus particles in a binary mixture, clustering is controlled by short-range excluded-volume interactions~\cite{Buttinoni2013}, whereas in the experiments performed by Palacci \emph{et al.}~\cite{Palacci2013}, the finite clusters were accounted for by the dominance of long-range phoretic interactions in that particular system.} A similar clustering phenomenon was observed for AgCl particles in the presence of UV light. In this situation, each particle responds to the ions gradients created by the other ones, combining diffusiophoretic and electrophoretic effects. Large particle schools spontaneously form, and can be reversibly destroyed by switching on and off the UV source~\cite{Ibele2009}.
\begin{figure}
\begin{center}
\includegraphics[width=\columnwidth]{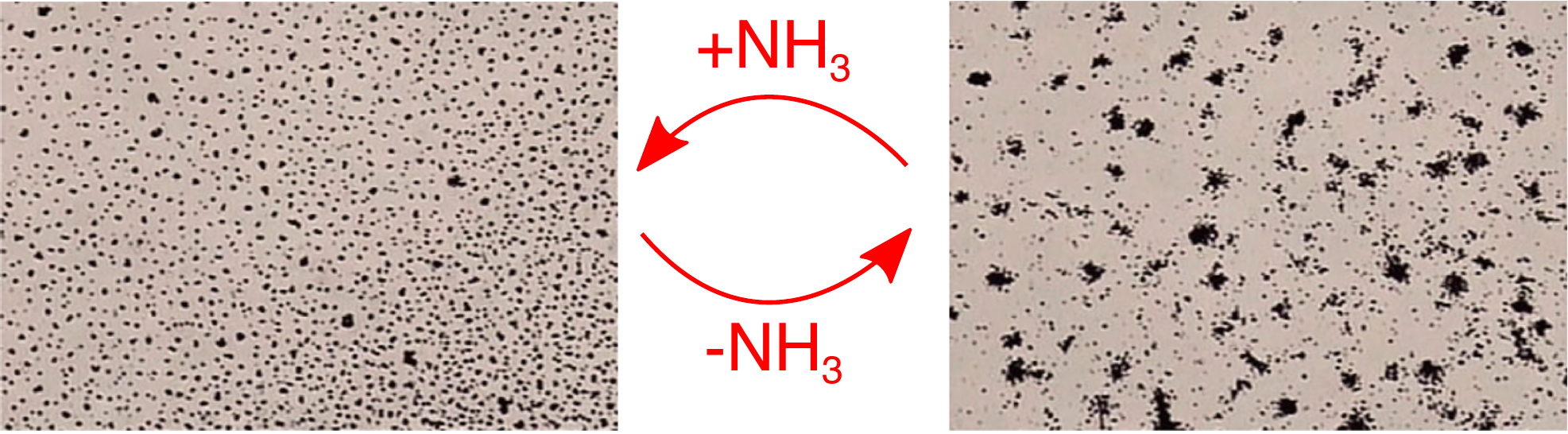}
\caption{Transition between exclusion and schooling behaviours based on self-diffusiophoresis in a suspension of Ag$_3$PO$_4$ microparticles~\cite{Duan2013}.}
\label{duan}
\end{center}
\end{figure}
The study of Ag$_3$PO$_4$ particles revealed the existence of two different regimes where the particles either exclude each other or form clusters. The reversible transition between these two states can be controlled by the intensity of UV light and by the addition of NH$_3$ in the solution (Fig. \ref{duan})~\cite{Duan2013}.

{Finally, acoustophoretic nanorods can form another class of complex patterns, which are quite specific, and cannot be easily related to the effects observed in assemblies of diffusiophoretic or electrophoretic spherical particles.} It was shown that they can assemble in long chains displaying axial rotation, or in very large circular patterns (up to 100 particle sizes)~\cite{Wang2012}.

\section{Statistical physics of interacting active colloids}

\subsection{Phenomenological and effective theories}

The experimental studies of active colloidal particles reveal the rich variety of their collective behaviours, which result from an interplay between the interactions that exist between them and the complexity of their individual dynamics. Because of the nonequilibrium nature of their behavior, the usual rules and theorems from equilibrium statistical physics do not apply, and there are no general principles governing their evolution.

For these reasons, the study of  assemblies of active particles from a statistical physics point of view is particularly challenging, and has motivated different approaches. Some of them rely on phenomenological descriptions: coarse-grained evolution equations for the density, polarisation and nematic fields can be obtained based on the symmetry properties of the physical system~\cite{Marchetti2013}. Detailed studies of these equations  gave an overview of the different behaviours emerging from active assemblies. In another class of theoretical approaches, the interactions between the active particles are taken into account in an effective way, for instance by assuming that the propulsion velocity is a function of the local density~\cite{Tailleur2008}. This efficient approach predicts the existence of a motility-induced phase separation, where the systems breaks between dense clusters and dilute regions.

Although these phenomenological or effective theories could be compared to some experimental situations, there are fewer theories that take into account explicitly the details of the interactions that are present in the systems. As it was emphasised in the previous section of this review, the evolution of the position of an individual phoretic  particle is controlled by the local inhomogeneities of the `phoretic field' (concentration, temperature...). In turn, the displacement of the active particle will modify the local structure of the field. This interplay between the phoretic particle and its environment becomes particularly complex when other active particles are present in the system. Each colloid then feels the presence of the others, and is affected by the other local field inhomogeneities. These interactions, that are typically long-ranged, could play a significant role in the emerging macroscopic  behaviours observed in assemblies of active particles. Moreover, most of the experiments presented above, as well as the realistic situation of interest, take place in a fluid solvent. At the considered length scales, viscous effects prevail, and the appropriate level of description is given by low Reynolds number hydrodynamics, in which the solvent mediates long-range interactions between the particles.

{In what follows, we review the different models (treated analytically or computationally) that take into account explicitly the interactions between active colloids. We aim to highlight the technical difficulties that arise in such approaches, and to discuss the emergence of the collective effects observed in experimental realisations. Due to size limitation of tutorial reviews, we will focus on our own contributions to this growing field of research. References to other significant advances can be found within the papers cited in this section.}

\subsection{Interacting thermophoretic colloids}

As a first example of the effect of long-range interactions on the collective dynamics of active colloids, we consider the case of thermophoretic particles. It was shown experimentally that silica colloidal particles half-coated with gold could self-propel when they are placed in a defocused laser beam~\cite{Jiang2010}. This self-generated temperature gradient is responsible for the individual propulsion of the colloids, with a velocity $\boldsymbol{v}$ that can be related to the intensity of the laser beam, the thermodiffusion coefficient $D_T$ and the thermal conductivity of the medium. In addition to this self-propulsion effect, each colloid will act as a local heat source, and will therefore creates an inhomogeneous temperature field $T(\boldsymbol{r})$, whose magnitude decreases as $1/r$. The drift velocity resulting from these temperature gradients is typically given by the relation $\boldsymbol{v} = -D_T\nabla T$. The superposition of the long-range  profiles created by the colloids will create a complex and dynamical temperature field that will be responsible for interactions between them, and that could ultimately lead to collective effects.

\begin{figure}
\begin{center}
\includegraphics[width=\columnwidth]{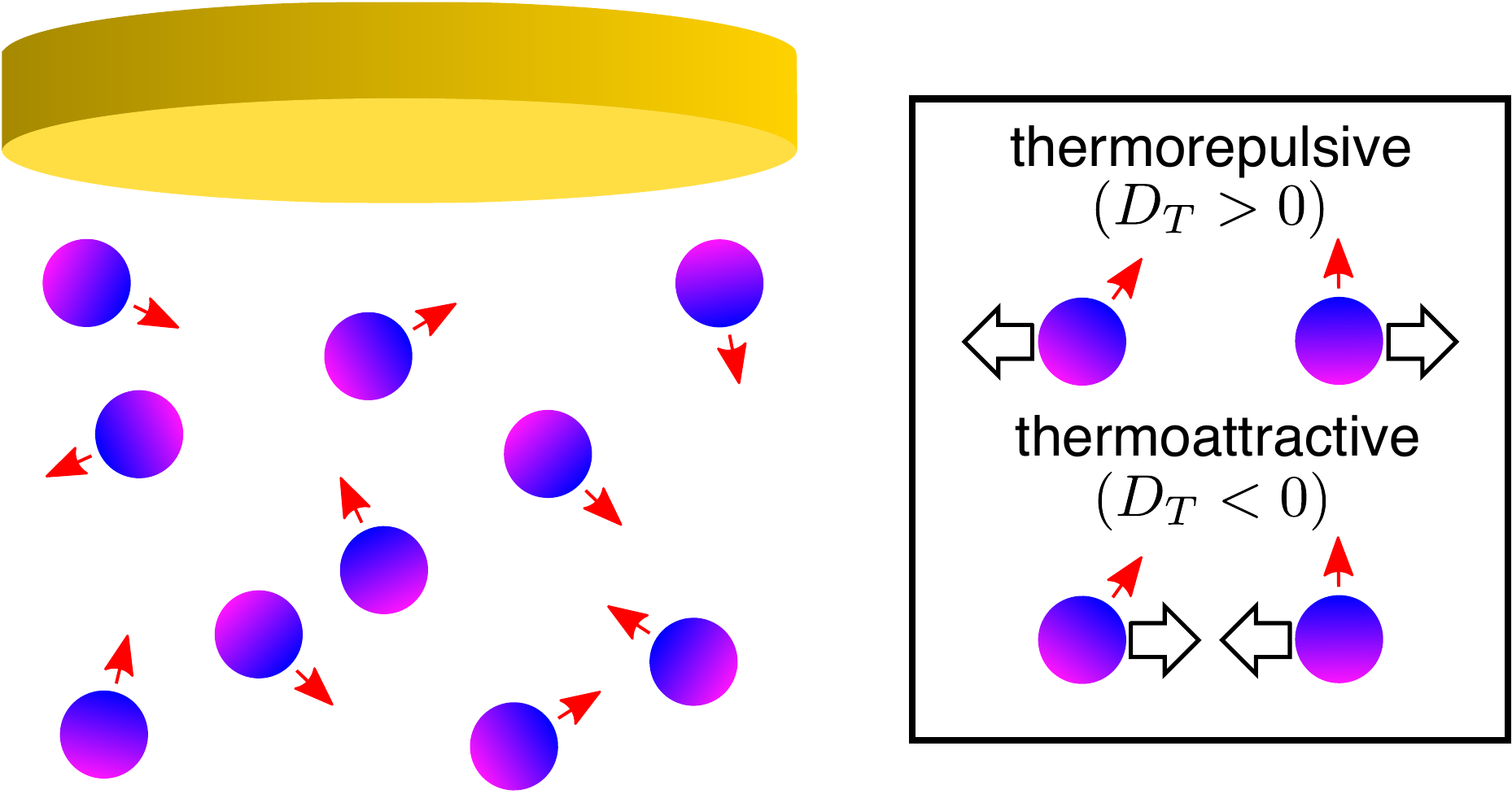}
\caption{{Particles with an asymmetric metallic coating can move because of thermophoretic effects when they are irradiated by a source of light, coming from the top in this simplified representation. Each particle moves in the direction of the small red arrow.  In addition to this self-propulsion mechanism, the colloids interact with each other because of the slowly decreasing temperature profiles they generate. Depending on the sign of thermodiffusion coefficient $D_T$, the interactions can be attractive or repulsive~\cite{Golestanian2012}.}}
\label{thermophoretic}
\end{center}
\end{figure}

In order to illustrate the technical difficulties that arise in the analytical description of this collective stochastic dynamics, we write the Langevin equations satisfied by the position $\boldsymbol{r}_i$ and orientation $\boldsymbol{n}_i$~\cite{Golestanian2012}:
\begin{eqnarray}
\dot{\boldsymbol{r}}_i & = & {V} \nn_i -D_T \nabla T(\rr_i,t) + {\bf \xi}_i, \\
\dot{\nn}_i & = & {\bf \eta}_i \times \nn_i ,
\end{eqnarray}
where ${\bf \xi}_i$ and ${\bf \eta}_i$ are Gaussian white noises, and where the hydrodynamic interactions are ignored for simplicity. The influence of the other colloids is taken into account in the temperature field $T$. At lower order in the multipole expansion, this field writes $T(\rr,t) = T_0+T_1\sum_j \frac{1}{|\rr-\rr_j(t)|}$. Defining a coarse-grained probability   ution $\mathcal{P}(\rr,\nn;t)=\moy{\sum_{i=1}^N \delta(\rr-\rr_i(t))\delta(\nn-\nn_i(t))}$, we can turn to a Fokker-Planck description of the stochastic dynamics, that is completed by a heat diffusion equation satisfied by the temperature field $T$. Determining the effective diffusion coefficient of the colloids requires averaging over the orientation $\nn$ in order to get the evolution equation of the density field $\rho(\rr;t)=\int_{\nn}\mathcal{P}(\rr,\nn;t)$. However, such an equation is not closed and involves higher-order moments, such as a polarisation field $\boldsymbol{p}(\rr;t)=\int_{\nn}\nn\mathcal{P}(\rr,\nn;t)$, whose evolution is determined in terms of second-order nematic fields and so on.

This hierarchy of equations, which formally resembles the classical Bogoliubov-Born-Green-Kirkwood-Yvon hierarchy, arises from the coupling between the orientational and translational degrees of freedom, and is a generic feature of the stochastic description of assemblies of active colloids. This hierarchy can be truncated by focusing on the limits of long time and of large length scales, which typically yields a closed set of equations satisfied by the density and polarisation fields. This closure approximation yields an expression for the effective diffusion coefficient of the colloids and a self-consistent equation for the temperature.

In the thermorepulsive case where the colloids repel each other ($D_T>0$, see Fig. \ref{thermophoretic}), the structure of the equations suggest an analogy with electrostatics, with typical length scales that play the role of the Bjerrum length and the Debye screening length. The system then exhibits a strong depletion effect under confinement. In the thermoattractive case where the colloids attract each other ($D_T<0$), the particles will be attracted towards the centre of the confinement box, suggesting an analogy with a gravitational system. In addition to these general analytical predictions, extensive numerical simulations of collections of attractive thermophoretic colloids revealed the formation of comet-like swarms moving towards the source of heat~\cite{Cohen2014}. The shape and the density fluctuations of the swarm could be investigated by modulating the coupling strength between the colloids.

This example shows that going beyond phenomenological descriptions of active systems, and taking into account the microscopic principles that rule the individual evolution of the colloids and their interactions could reveal a very rich collection of collective behaviours, that would be missed in an effective description, and that may play a significant role in experimental systems.

\subsection{Interactions mediated by chemical fields}

\subsubsection{Self-assembled active molecules}

We now turn to the case of colloids that interact \emph{via} chemical fields. As it was detailed for the case of an individual colloid, the motion of diffusiophoretic particles is permitted by the existence of inhomogeneities in the concentration of solute molecules with which the colloid interacts. Let us consider symmetric colloidal particles whose surface acts as a sink or a source of solute molecules. The concentration profile is asymptotically given by $C(r)\sim\alpha R^2/(D r)$, where $D$ is the diffusion coefficient of the solute molecules and $\alpha$ is the surface activity, whose sign determines whether chemicals are consumed or produced by the colloid. The interaction of the colloid with the chemical field is determined by the surface mobility $\mu$. Such a symmetric colloid, when it is considered alone, does not have a directed motion since no symmetry-breaking occurs in the system. However, when several colloids with different values for the surface activities and mobilities are brought together, the resulting phoretic interactions could lead to net motion.

\begin{figure}
\begin{center}
\includegraphics[width=\columnwidth]{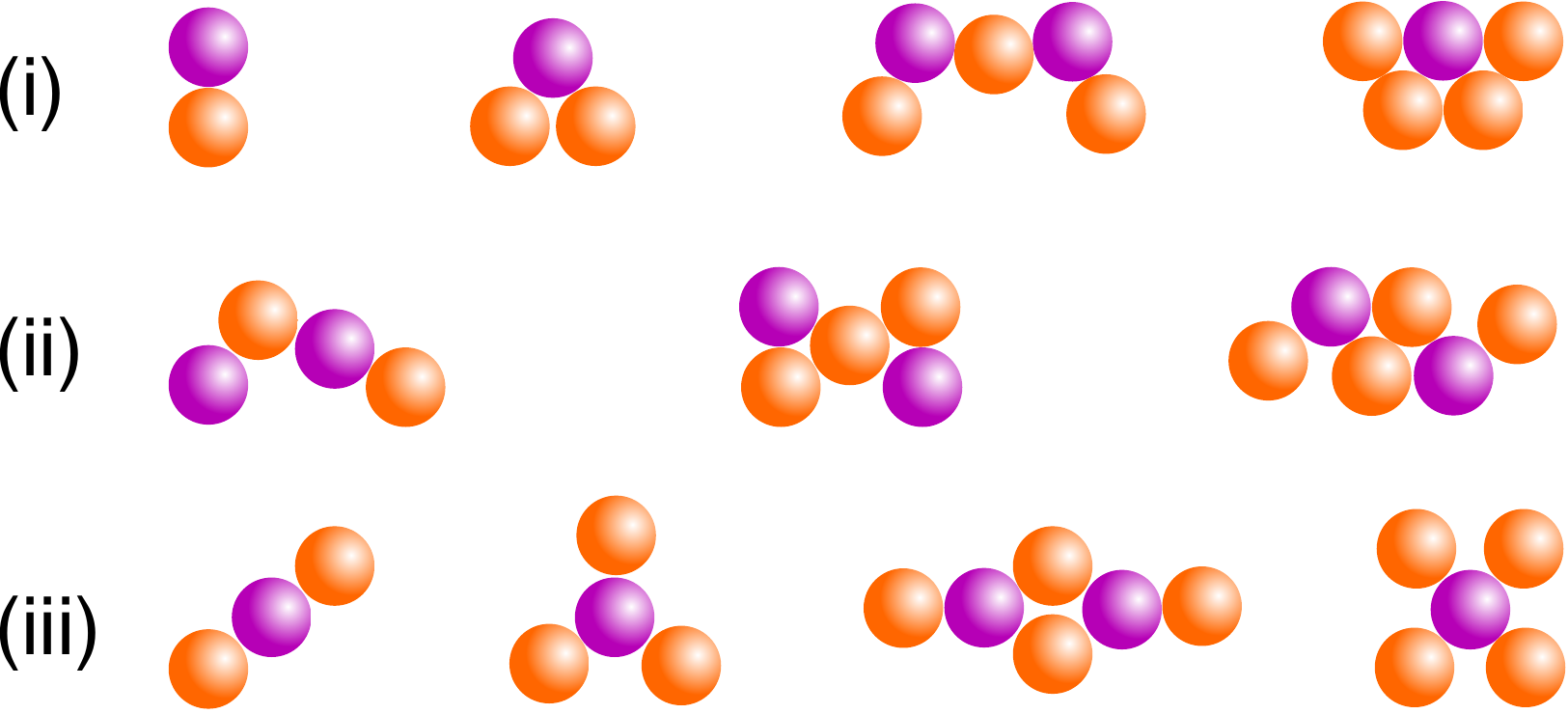}
\caption{Examples of self-assembled molecules made of two types of colloids with different surface mobilities and activities~\cite{Soto2014}. The symmetry properties of these small clusters give rise to different types of motion, namely self-propulsion (i) or self-rotation (ii). Some inert molecules are also formed (iii).}
\label{active_molecules}
\end{center}
\end{figure}

It was recently shown with Brownian dynamics simulations that assemblies made of two types of particles in dilute conditions could lead to the formation of stable active `molecules'~\cite{Soto2014}. These objects are made of combinations of the two types of colloids (see Fig. \ref{active_molecules} for examples). Depending on the geometries of these clusters, they can either be inert (with no net motion) or exhibit net self-propulsion or self-rotation. A more thorough investigation of the dynamics of these molecules revealed that they could exhibit spontaneous oscillations between different conformations or run-and-tumble motion~\cite{Soto2015}. These observations show that phoretic interactions can lead to self-organization of active colloids into complex clusters, and highlights once again their significance.

\subsubsection{Collective behaviour of active colloids in a chemical gradient}

The examples presented so far could be classified into two different categories: the field gradient was either generated by the active particle itself, or was imposed externally by the experimental conditions. However, the situation where an asymmetric particle moves in a macroscopic gradient is a combination of the two cases studied above, and needs to be considered in greater details. This could also describe situations encountered in biological systems, either at the intracellular or cellular levels, where cells or macromolecular complexes with built-in asymmetries are able to perform complex functions in strongly inhomogeneous environments.

\begin{figure}
\begin{center}
\includegraphics[width=\columnwidth]{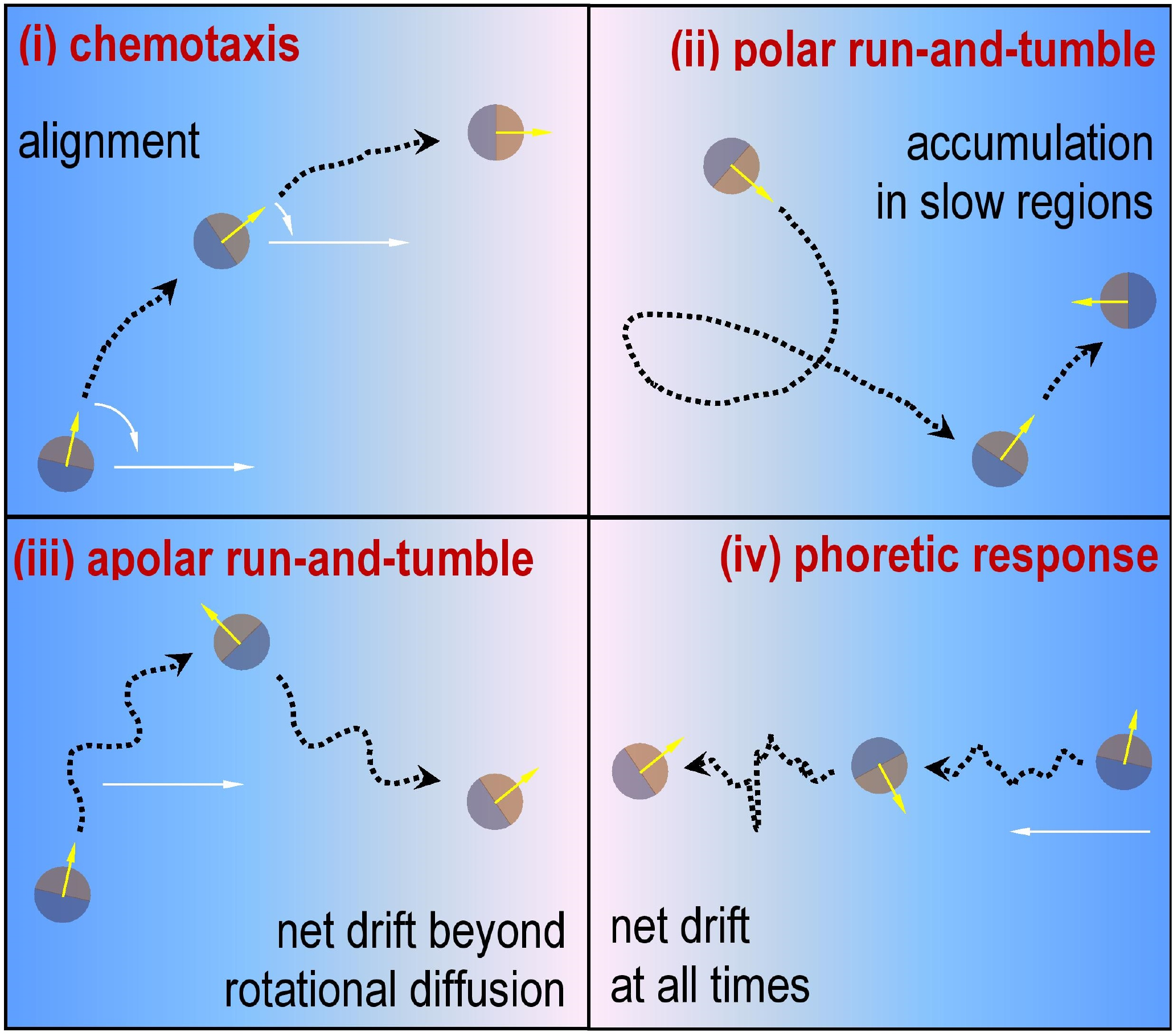}
\caption{The different responses of a self-diffusiophoretic active colloid placed in a gradient of chemical substrate~\cite{Saha2014}.}
\label{chemotaxis}
\end{center}
\end{figure}

Let us consider the particular case of an asymmetric self-diffusiophoretic colloid placed in a macroscopic gradient of a chemical reactant of concentration $s$. The transformation of this reactant into product molecules occurs at the surface of the colloid, which is asymmetrically coated with a catalyst. Without considering any collective effects, the individual behaviour of the colloid is itself quite complicated, as it results from the interplay between the phoretic and the chemotactic responses to its environment. The centre-of-mass velocity and the angular velocity of the particle can take the following generic forms:
\begin{eqnarray}
{\bf \omega} & = & \Phi_0 \nn \times \nabla s \label{saha1}, \\
\boldsymbol{v} & = & V_0 \nn - \alpha_0 \nabla s -\alpha_1 \nn\nn \cdot \nabla s \label{saha2}.
\end{eqnarray}
Each of the four parameters from these equations ($\Phi_0$, $V_0$, $\alpha_0$ and $\alpha_1$) characterises a different response of the colloid to the gradients present in Eqs. \ref{saha1} and \ref{saha2}, and a corresponding sketch is presented in Fig. \ref{chemotaxis}: (i) $\Phi_0$ represents the angular drift of the colloid, which is related to its chemotactic response: the orientation of the particle tends to align parallel or antiparallel to the local chemical gradient; (ii) the velocity $V_0$ is controlled by the reaction rate, so that a polar run-and-tumble mechanism would tend to attract the colloid in slow regions; (iii) an apolar run-and-tumble mechanism driven by the gradient  brings the colloid in the fast regions (controlled by $\alpha_1$); (iv) phoretic response (controlled by $\alpha_0$) is responsible for a net drift of the position of the centre of mass of the colloid, independently of its polarity.

These four parameters can be explicitly related to the surface mobility $\mu(\rr_\text{s})$ and activity $\alpha(\rr_\text{s})$. Assuming axial symmetry of the colloid coating, these quantities can usually be written as spherical harmonics expansions, that are easy to handle mathematically when suitably truncated~\cite{Golestanian2007}.

The presence of multiple active colloids gives rise to effective interactions which are mediated by the reactant and product concentration fields, themselves affected by the activity of each individual, and that are typically long-ranged ($\sim 1/r$). The interplay between the chemotactic and phoretic contributions that control the displacement of each colloid may give rise to a complex phase diagram. In order to get a more quantitative insight in these collective effects, the stochastic dynamics of the system is described with a set of Langevin equations satisfied by the position and orientation of each active particle. This Langevin description can be turned into a single Fokker-Planck equation for the joint coarse-grained probability density of position and orientation. Like in the case of interacting thermophoretic colloids, the long-range phoretic interactions are responsible for a coupling between the translational and orientational degrees of freedom, which makes the problem particularly difficult to treat. A moment expansion with respect to the orientational variable and a closure approximation allows one to obtain a closed set of equations for the density and polarisation fields. It is finally possible to sketch the phase diagram of the active assembly in a simplified two-dimensional parameter space, where one parameter describes the chemotactic response and the other one the phoretic response of the colloids~\cite{Saha2014}.

Assuming that the formation of product molecules obeys a Michaelis-Menten-like kinetics (linear in the reactant concentration for small $s$ and saturated for large $s$), two different concentration regimes can be identified. When the fuel concentration is small, the activity is diffusion-limited, which means that the typical reaction time is much faster than the time taken by substrate molecules to diffuse up to the surface of an active colloid. In this limit, the chemical fields are screened, and the phase diagram reveals the emergence of large number fluctuations and clumping instabilities. In the saturation regime where the reactant concentration is very high and where the reaction rate is the limiting timescale of the chemical activity, the chemical fields are not screened anymore, and their long-ranged character gives rise to different collective effects, among which a Jeans instability, the formation of asters, and the emergence of oscillating patterns.

In this model, the solute molecules are consumed and produced at the surfaces of the colloids only, and there are no chemical reactions occurring in the bulk. However, in order to model accurately active transport in biological systems, it is necessary to take into account the underlying chemical activity and the nonequilibrium character of the environment of the particles. Considering a reaction-diffusion equation coupled to the dynamics of the phoretic particles, Banigan and Marko showed analytically and numerically how the background activity can modulate the effective interactions between the colloids, and affect the self-organisation of the system~\cite{Banigan2016}.

\subsection{The role of hydrodynamics}

The analytical and numerical study of assemblies of active colloids usually ignore the presence of the liquid solvent in which the particles evolve, whose description is particularly complex.  However, given the typical length scales (micrometer) and velocities (tens to hundreds of micrometers per second) of self-propelled active colloids, the viscous effects are expected to prevail, and the appropriate level of description is actually given by the laws of low Reynolds number hydrodynamics. Such a hydrodynamic description of the system would allow one to take into account the continuous energy dissipation mediated by the viscous liquid, as well as the hydrodynamic interactions, who typically decrease slowly (as the inverse of the distance) and who could also play a significant role in the near-field limit. In particular, it would be important to understand the role played by hydrodynamic interactions in the emergence of the macroscopic effects observed experimentally, such as the formation of large clusters separated by gas-like regions.

A first approach to this question was provided by Z\"ottl and Stark~\cite{Zottl2014}, who investigated numerically the evolution of assemblies of squirmers (particles that self-propel by generating an axisymmetric slip velocity) in a confined quasi-2D system, and taking into account the hydrodynamic interactions. They observe a transition to clustering, and they show that this transition is controlled by the hydrodynamics near field, that prevail when the particles collide frequently. It was subsequently showed  by simulating hydrodynamic squirmers in an unbounded fluid that the presence of hydrodynamic interactions could actually suppress the phase separation predicted by the usual motility-induced mechanisms~\cite{Matas-Navarro2014}. The discrepancy between these two set of numerical results can be accounted for by the fact that the geometrical confinement actually screens the hydrodynamic interactions, with a screening length of the order of the thickness of the quasi-2D slab. The comparison of these results then allows one to distinguish the role of the far-field effects from that of the near-field effects.

\section{Conclusion}

In the context of phoretic motion, `fuelling' means driving steadily the system to a nonequilibrium state, by having intrinsic symmetry breaking mechanisms that can convert a uniform free energy source into directed motion. This generic idea lead to the theoretical study and to the experimental design of many self-propelled particles, which rely on different microscopic mechanisms, but which are formally equivalent. In addition to their surprising individual properties, the collective behaviour of active colloids, that give rise to unexpected macroscopic effects, has motivated many studies. As it was described in the present review, this field of research has been particularly productive during the past years and has brought together the effort of different scientific communities, who are now facing new challenges.

From an experimental perspective, controlling the motion of self-propelled colloids could open the way to numerous applications, in particular in the biomedical sciences. More complex functionalisations of the surface of the colloid could allow them to perform specific functions, such as carrying smaller particles to a target, following a given trajectory, or even reproducing some intracellular functions such as replication or gradient-sensing. In this perspective, recent experimental breakthroughs showed how synthetic particles could use the chemical activity of biomolecules as a fuel to move in a directed manner, and to respond to environmental inhomogeneities~\cite{Dey2015}.

Active colloids, and in particular their collective behaviour, also raise a number of theoretical issues. Their description falls into the scope of non-equilibrium statistical physics, who does not rely on generic principles, so that each system needs to be studied with its specific details. Moreover, the long-ranged interactions between the active particles make any theoretical description particularly difficult, and little is known about the effect of phoretic and hydrodynamic interactions. This opens the way to new approaches, that rely on extensive numerical simulations and on the development of appropriate mathematical tools.

\section*{Acknowledgments}

We acknowledge financial support from the US National Science Foundation under MRSEC Grant No. DMR-1420620.






\providecommand*{\mcitethebibliography}{\thebibliography}
\csname @ifundefined\endcsname{endmcitethebibliography}
{\let\endmcitethebibliography\endthebibliography}{}


\begin{mcitethebibliography}{50}
\providecommand*{\natexlab}[1]{#1}
\providecommand*{\mciteSetBstSublistMode}[1]{}
\providecommand*{\mciteSetBstMaxWidthForm}[2]{}
\providecommand*{\mciteBstWouldAddEndPuncttrue}
  {\def\EndOfBibitem{\unskip.}}
\providecommand*{\mciteBstWouldAddEndPunctfalse}
  {\let\EndOfBibitem\relax}
\providecommand*{\mciteSetBstMidEndSepPunct}[3]{}
\providecommand*{\mciteSetBstSublistLabelBeginEnd}[3]{}
\providecommand*{\EndOfBibitem}{}
\mciteSetBstSublistMode{f}
\mciteSetBstMaxWidthForm{subitem}
{(\emph{\alph{mcitesubitemcount}})}
\mciteSetBstSublistLabelBeginEnd{\mcitemaxwidthsubitemform\space}
{\relax}{\relax}

\bibitem[Bechinger \emph{et~al.}(2016)Bechinger, {Di Leonardo}, L{\"{o}}wen,
  Reichhardt, Volpe, and Volpe]{Bechinger2016}
C.~Bechinger, R.~{Di Leonardo}, H.~L{\"{o}}wen, C.~Reichhardt, G.~Volpe and
  G.~Volpe, \emph{Rev. Mod. Phys.}, 2016, \textbf{88}, 045006\relax
\mciteBstWouldAddEndPuncttrue
\mciteSetBstMidEndSepPunct{\mcitedefaultmidpunct}
{\mcitedefaultendpunct}{\mcitedefaultseppunct}\relax
\EndOfBibitem
\bibitem[Wong \emph{et~al.}(2016)Wong, Dey, and Sen]{Wong2016}
F.~Wong, K.~K. Dey and A.~Sen, \emph{Ann. Rev. Mater. Res.}, 2016, \textbf{46},
  407--432\relax
\mciteBstWouldAddEndPuncttrue
\mciteSetBstMidEndSepPunct{\mcitedefaultmidpunct}
{\mcitedefaultendpunct}{\mcitedefaultseppunct}\relax
\EndOfBibitem
\bibitem[Anderson(1989)]{Anderson1989}
J.~Anderson, \emph{Annual Review of Fluid Mechanics}, 1989, \textbf{21},
  61--99\relax
\mciteBstWouldAddEndPuncttrue
\mciteSetBstMidEndSepPunct{\mcitedefaultmidpunct}
{\mcitedefaultendpunct}{\mcitedefaultseppunct}\relax
\EndOfBibitem
\bibitem[Golestanian \emph{et~al.}(2005)Golestanian, Liverpool, and
  Ajdari]{Golestanian2005}
R.~Golestanian, T.~B. Liverpool and A.~Ajdari, \emph{Phys. Rev. Lett.}, 2005,
  \textbf{94}, 220801\relax
\mciteBstWouldAddEndPuncttrue
\mciteSetBstMidEndSepPunct{\mcitedefaultmidpunct}
{\mcitedefaultendpunct}{\mcitedefaultseppunct}\relax
\EndOfBibitem
\bibitem[Nardi \emph{et~al.}(1999)Nardi, Bruinsma, and Sackmann]{Nardi1999a}
J.~Nardi, R.~Bruinsma and E.~Sackmann, \emph{Physical Review Letters}, 1999,
  \textbf{82}, 5168--5171\relax
\mciteBstWouldAddEndPuncttrue
\mciteSetBstMidEndSepPunct{\mcitedefaultmidpunct}
{\mcitedefaultendpunct}{\mcitedefaultseppunct}\relax
\EndOfBibitem
\bibitem[Michelin and Lauga(2014)]{Michelin2014}
S.~Michelin and E.~Lauga, \emph{J. Fluid Mech}, 2014, \textbf{747},
  572--604\relax
\mciteBstWouldAddEndPuncttrue
\mciteSetBstMidEndSepPunct{\mcitedefaultmidpunct}
{\mcitedefaultendpunct}{\mcitedefaultseppunct}\relax
\EndOfBibitem
\bibitem[Ajdari and Bocquet(2006)]{Ajdari2006}
A.~Ajdari and L.~Bocquet, \emph{Phys. Rev. Lett.}, 2006, \textbf{96},
  186102\relax
\mciteBstWouldAddEndPuncttrue
\mciteSetBstMidEndSepPunct{\mcitedefaultmidpunct}
{\mcitedefaultendpunct}{\mcitedefaultseppunct}\relax
\EndOfBibitem
\bibitem[Golestanian \emph{et~al.}(2007)Golestanian, Liverpool, and
  Ajdari]{Golestanian2007}
R.~Golestanian, T.~B. Liverpool and A.~Ajdari, \emph{New Journal of Physics},
  2007, \textbf{9}, 126\relax
\mciteBstWouldAddEndPuncttrue
\mciteSetBstMidEndSepPunct{\mcitedefaultmidpunct}
{\mcitedefaultendpunct}{\mcitedefaultseppunct}\relax
\EndOfBibitem
\bibitem[Howse \emph{et~al.}(2007)Howse, Jones, Ryan, Gough, Vafabakhsh, and
  Golestanian]{Howse2007}
J.~R. Howse, R.~A.~L. Jones, A.~J. Ryan, T.~Gough, R.~Vafabakhsh and
  R.~Golestanian, \emph{Phys. Rev. Lett.}, 2007, \textbf{99}, 048102\relax
\mciteBstWouldAddEndPuncttrue
\mciteSetBstMidEndSepPunct{\mcitedefaultmidpunct}
{\mcitedefaultendpunct}{\mcitedefaultseppunct}\relax
\EndOfBibitem
\bibitem[Golestanian(2009)]{Golestanian2009}
R.~Golestanian, \emph{Phys. Rev. Lett.}, 2009, \textbf{102}, 188305\relax
\mciteBstWouldAddEndPuncttrue
\mciteSetBstMidEndSepPunct{\mcitedefaultmidpunct}
{\mcitedefaultendpunct}{\mcitedefaultseppunct}\relax
\EndOfBibitem
\bibitem[Paxton \emph{et~al.}(2004)Paxton, Kistler, Olmeda, Sen, {St. Angelo},
  Cao, Mallouk, Lammert, and Crespi]{Paxton2004}
W.~F. Paxton, K.~C. Kistler, C.~C. Olmeda, A.~Sen, S.~K. {St. Angelo}, Y.~Cao,
  T.~E. Mallouk, P.~E. Lammert and V.~H. Crespi, \emph{Journal of the American
  Chemical Society}, 2004, \textbf{126}, 13424--13431\relax
\mciteBstWouldAddEndPuncttrue
\mciteSetBstMidEndSepPunct{\mcitedefaultmidpunct}
{\mcitedefaultendpunct}{\mcitedefaultseppunct}\relax
\EndOfBibitem
\bibitem[Paxton \emph{et~al.}(2006)Paxton, Baker, Kline, Wang, Mallouk, and
  Sen]{Paxton2006}
W.~F. Paxton, P.~T. Baker, T.~R. Kline, Y.~Wang, T.~E. Mallouk and A.~Sen,
  \emph{Journal of the American Chemical Society}, 2006, \textbf{128},
  14881--14888\relax
\mciteBstWouldAddEndPuncttrue
\mciteSetBstMidEndSepPunct{\mcitedefaultmidpunct}
{\mcitedefaultendpunct}{\mcitedefaultseppunct}\relax
\EndOfBibitem
\bibitem[Wang \emph{et~al.}(2013)Wang, Duan, Sen, and Mallouk]{Wang2013}
W.~Wang, W.~Duan, A.~Sen and T.~E. Mallouk, \emph{Proceedings of the National
  Academy of Sciences of the United States of America}, 2013, \textbf{110},
  17744--9\relax
\mciteBstWouldAddEndPuncttrue
\mciteSetBstMidEndSepPunct{\mcitedefaultmidpunct}
{\mcitedefaultendpunct}{\mcitedefaultseppunct}\relax
\EndOfBibitem
\bibitem[Hong \emph{et~al.}(2007)Hong, Blackman, Kopp, Sen, and
  Velegol]{Hong2007}
Y.~Hong, N.~M.~K. Blackman, N.~D. Kopp, A.~Sen and D.~Velegol, \emph{Phys. Rev.
  Lett.}, 2007, \textbf{99}, 178103\relax
\mciteBstWouldAddEndPuncttrue
\mciteSetBstMidEndSepPunct{\mcitedefaultmidpunct}
{\mcitedefaultendpunct}{\mcitedefaultseppunct}\relax
\EndOfBibitem
\bibitem[Catchmark \emph{et~al.}(2005)Catchmark, Subramanian, and
  Sen]{Catchmark2005}
J.~M. Catchmark, S.~Subramanian and A.~Sen, \emph{Small}, 2005, \textbf{1},
  202--206\relax
\mciteBstWouldAddEndPuncttrue
\mciteSetBstMidEndSepPunct{\mcitedefaultmidpunct}
{\mcitedefaultendpunct}{\mcitedefaultseppunct}\relax
\EndOfBibitem
\bibitem[Ebbens \emph{et~al.}(2014)Ebbens, Gregory, Dunderdale, Howse, Ibrahim,
  Liverpool, and Golestanian]{Ebbens2014}
S.~Ebbens, D.~A. Gregory, G.~Dunderdale, J.~R. Howse, Y.~Ibrahim, T.~B.
  Liverpool and R.~Golestanian, \emph{Europhysics Letters (EPL)}, 2014,
  \textbf{106}, 58003\relax
\mciteBstWouldAddEndPuncttrue
\mciteSetBstMidEndSepPunct{\mcitedefaultmidpunct}
{\mcitedefaultendpunct}{\mcitedefaultseppunct}\relax
\EndOfBibitem
\bibitem[Das \emph{et~al.}(2015)Das, Garg, Campbell, Howse, Sen, Velegol,
  Golestanian, and Ebbens]{Das2015}
S.~Das, A.~Garg, A.~I. Campbell, J.~Howse, A.~Sen, D.~Velegol, R.~Golestanian
  and S.~J. Ebbens, \emph{Nature communications}, 2015, \textbf{6}, 8999\relax
\mciteBstWouldAddEndPuncttrue
\mciteSetBstMidEndSepPunct{\mcitedefaultmidpunct}
{\mcitedefaultendpunct}{\mcitedefaultseppunct}\relax
\EndOfBibitem
\bibitem[Simmchen \emph{et~al.}(2016)Simmchen, Katuri, Uspal, Popescu,
  Tasinkevych, and S{\'{a}}nchez]{Simmchen2016}
J.~Simmchen, J.~Katuri, W.~E. Uspal, M.~N. Popescu, M.~Tasinkevych and
  S.~S{\'{a}}nchez, \emph{Nature Communications}, 2016, \textbf{7}, 10598\relax
\mciteBstWouldAddEndPuncttrue
\mciteSetBstMidEndSepPunct{\mcitedefaultmidpunct}
{\mcitedefaultendpunct}{\mcitedefaultseppunct}\relax
\EndOfBibitem
\bibitem[Palacci \emph{et~al.}(2013)Palacci, Sacanna, Steinberg, Pine, and
  Chaikin]{Palacci2013}
J.~Palacci, S.~Sacanna, A.~P. Steinberg, D.~J. Pine and P.~M. Chaikin,
  \emph{Science}, 2013, \textbf{339}, 936--340\relax
\mciteBstWouldAddEndPuncttrue
\mciteSetBstMidEndSepPunct{\mcitedefaultmidpunct}
{\mcitedefaultendpunct}{\mcitedefaultseppunct}\relax
\EndOfBibitem
\bibitem[Ibele \emph{et~al.}(2009)Ibele, Mallouk, and Sen]{Ibele2009}
M.~Ibele, T.~E. Mallouk and A.~Sen, \emph{Angewandte Chemie - International
  Edition}, 2009, \textbf{48}, 3308--3312\relax
\mciteBstWouldAddEndPuncttrue
\mciteSetBstMidEndSepPunct{\mcitedefaultmidpunct}
{\mcitedefaultendpunct}{\mcitedefaultseppunct}\relax
\EndOfBibitem
\bibitem[Ibele \emph{et~al.}(2010)Ibele, Lammert, Crespi, and Sen]{Ibele2010c}
M.~E. Ibele, P.~E. Lammert, V.~H. Crespi and A.~Sen, \emph{ACS Nano}, 2010,
  \textbf{4}, 4845--4851\relax
\mciteBstWouldAddEndPuncttrue
\mciteSetBstMidEndSepPunct{\mcitedefaultmidpunct}
{\mcitedefaultendpunct}{\mcitedefaultseppunct}\relax
\EndOfBibitem
\bibitem[Hong \emph{et~al.}(2010)Hong, Diaz, C{\'{o}}rdova-Fteueroa, and
  Sen]{Hong2010a}
Y.~Hong, M.~Diaz, U.~M. C{\'{o}}rdova-Fteueroa and A.~Sen, \emph{Advanced
  Functional Materials}, 2010, \textbf{20}, 1568--1576\relax
\mciteBstWouldAddEndPuncttrue
\mciteSetBstMidEndSepPunct{\mcitedefaultmidpunct}
{\mcitedefaultendpunct}{\mcitedefaultseppunct}\relax
\EndOfBibitem
\bibitem[Duan \emph{et~al.}(2013)Duan, Liu, and Sen]{Duan2013}
W.~Duan, R.~Liu and A.~Sen, \emph{Journal of the American Chemical Society},
  2013, \textbf{135}, 1280--1283\relax
\mciteBstWouldAddEndPuncttrue
\mciteSetBstMidEndSepPunct{\mcitedefaultmidpunct}
{\mcitedefaultendpunct}{\mcitedefaultseppunct}\relax
\EndOfBibitem
\bibitem[Rings \emph{et~al.}(2010)Rings, Schachoff, Selmke, Cichos, and
  Kroy]{Rings2010}
D.~Rings, R.~Schachoff, M.~Selmke, F.~Cichos and K.~Kroy, \emph{Phys. Rev.
  Lett.}, 2010, \textbf{105}, 090604\relax
\mciteBstWouldAddEndPuncttrue
\mciteSetBstMidEndSepPunct{\mcitedefaultmidpunct}
{\mcitedefaultendpunct}{\mcitedefaultseppunct}\relax
\EndOfBibitem
\bibitem[Volpe \emph{et~al.}(2011)Volpe, Buttinoni, Vogt, K{\"{u}}mmerer,
  Bechinger, Kuemmerer, and Bechinger]{Volpe2011}
G.~Volpe, I.~Buttinoni, D.~Vogt, H.-J. K{\"{u}}mmerer, C.~Bechinger, H.-J.
  Kuemmerer and C.~Bechinger, \emph{Soft Matter}, 2011, \textbf{7}, 8810\relax
\mciteBstWouldAddEndPuncttrue
\mciteSetBstMidEndSepPunct{\mcitedefaultmidpunct}
{\mcitedefaultendpunct}{\mcitedefaultseppunct}\relax
\EndOfBibitem
\bibitem[Buttinoni \emph{et~al.}(2013)Buttinoni, Bialk{\'{e}}, K{\"{u}}mmel,
  L{\"{o}}wen, Bechinger, and Speck]{Buttinoni2013}
I.~Buttinoni, J.~Bialk{\'{e}}, F.~K{\"{u}}mmel, H.~L{\"{o}}wen, C.~Bechinger
  and T.~Speck, \emph{Physical Review Letters}, 2013, \textbf{110},
  238301\relax
\mciteBstWouldAddEndPuncttrue
\mciteSetBstMidEndSepPunct{\mcitedefaultmidpunct}
{\mcitedefaultendpunct}{\mcitedefaultseppunct}\relax
\EndOfBibitem
\bibitem[W{\"{u}}rger(2015)]{Wurger2015}
A.~W{\"{u}}rger, \emph{Physical Review Letters}, 2015, \textbf{115},
  188304\relax
\mciteBstWouldAddEndPuncttrue
\mciteSetBstMidEndSepPunct{\mcitedefaultmidpunct}
{\mcitedefaultendpunct}{\mcitedefaultseppunct}\relax
\EndOfBibitem
\bibitem[Thutupalli \emph{et~al.}(2011)Thutupalli, Seemann, and
  Herminghaus]{Thutupalli2011}
S.~Thutupalli, R.~Seemann and S.~Herminghaus, \emph{New Journal of Physics},
  2011, \textbf{13}, 073021\relax
\mciteBstWouldAddEndPuncttrue
\mciteSetBstMidEndSepPunct{\mcitedefaultmidpunct}
{\mcitedefaultendpunct}{\mcitedefaultseppunct}\relax
\EndOfBibitem
\bibitem[Izri \emph{et~al.}(2014)Izri, {Van Der Linden}, Michelin, and
  Dauchot]{Izri2014}
Z.~Izri, M.~N. {Van Der Linden}, S.~Michelin and O.~Dauchot, \emph{Physical
  Review Letters}, 2014, \textbf{113}, 248302\relax
\mciteBstWouldAddEndPuncttrue
\mciteSetBstMidEndSepPunct{\mcitedefaultmidpunct}
{\mcitedefaultendpunct}{\mcitedefaultseppunct}\relax
\EndOfBibitem
\bibitem[Wang \emph{et~al.}(2012)Wang, Castro, Hoyos, and Mallouk]{Wang2012}
W.~Wang, L.~A. Castro, M.~Hoyos and T.~E. Mallouk, \emph{ACS Nano}, 2012,
  \textbf{6}, 6122--6132\relax
\mciteBstWouldAddEndPuncttrue
\mciteSetBstMidEndSepPunct{\mcitedefaultmidpunct}
{\mcitedefaultendpunct}{\mcitedefaultseppunct}\relax
\EndOfBibitem
\bibitem[Wang \emph{et~al.}(2014)Wang, Li, Mair, Ahmed, Huang, and
  Mallouk]{Wang2014}
W.~Wang, S.~Li, L.~Mair, S.~Ahmed, T.~J. Huang and T.~E. Mallouk,
  \emph{Angewandte Chemie - International Edition}, 2014, \textbf{53},
  3201--3204\relax
\mciteBstWouldAddEndPuncttrue
\mciteSetBstMidEndSepPunct{\mcitedefaultmidpunct}
{\mcitedefaultendpunct}{\mcitedefaultseppunct}\relax
\EndOfBibitem
\bibitem[Bricard \emph{et~al.}(2013)Bricard, Caussin, Desreumaux, Dauchot, and
  Bartolo]{Bricard2013}
A.~Bricard, J.-B. Caussin, N.~Desreumaux, O.~Dauchot and D.~Bartolo,
  \emph{Nature}, 2013, \textbf{503}, 95\relax
\mciteBstWouldAddEndPuncttrue
\mciteSetBstMidEndSepPunct{\mcitedefaultmidpunct}
{\mcitedefaultendpunct}{\mcitedefaultseppunct}\relax
\EndOfBibitem
\bibitem[Purcell(1977)]{Purcell1977}
E.~M. Purcell, \emph{Am. J. Phys.}, 1977, \textbf{45}, 3\relax
\mciteBstWouldAddEndPuncttrue
\mciteSetBstMidEndSepPunct{\mcitedefaultmidpunct}
{\mcitedefaultendpunct}{\mcitedefaultseppunct}\relax
\EndOfBibitem
\bibitem[Najafi and Golestanian(2004)]{Najafi2004}
A.~Najafi and R.~Golestanian, \emph{Phys. Rev. E}, 2004, \textbf{69},
  062901\relax
\mciteBstWouldAddEndPuncttrue
\mciteSetBstMidEndSepPunct{\mcitedefaultmidpunct}
{\mcitedefaultendpunct}{\mcitedefaultseppunct}\relax
\EndOfBibitem
\bibitem[Dreyfus \emph{et~al.}(2005)Dreyfus, Baudry, Roper, Fermigier, Stone,
  and Bibette]{Dreyfus2005}
R.~Dreyfus, J.~Baudry, M.~L. Roper, M.~Fermigier, H.~a. Stone and J.~Bibette,
  \emph{Nature}, 2005, \textbf{437}, 862--865\relax
\mciteBstWouldAddEndPuncttrue
\mciteSetBstMidEndSepPunct{\mcitedefaultmidpunct}
{\mcitedefaultendpunct}{\mcitedefaultseppunct}\relax
\EndOfBibitem
\bibitem[Tierno \emph{et~al.}(2008)Tierno, Golestanian, Pagonabarraga, and
  Sagu{\'{e}}s]{Tierno2008}
P.~Tierno, R.~Golestanian, I.~Pagonabarraga and F.~Sagu{\'{e}}s, \emph{Physical
  Review Letters}, 2008, \textbf{101}, 218304\relax
\mciteBstWouldAddEndPuncttrue
\mciteSetBstMidEndSepPunct{\mcitedefaultmidpunct}
{\mcitedefaultendpunct}{\mcitedefaultseppunct}\relax
\EndOfBibitem
\bibitem[Ghosh and Fischer(2009)]{Ghost2009}
A.~Ghosh and P.~Fischer, \emph{Nano Letters}, 2009, \textbf{9},
  2243--2245\relax
\mciteBstWouldAddEndPuncttrue
\mciteSetBstMidEndSepPunct{\mcitedefaultmidpunct}
{\mcitedefaultendpunct}{\mcitedefaultseppunct}\relax
\EndOfBibitem
\bibitem[Theurkauff \emph{et~al.}(2012)Theurkauff, Cottin-Bizonne, Palacci,
  Ybert, and Bocquet]{Theurkauff2012}
I.~Theurkauff, C.~Cottin-Bizonne, J.~Palacci, C.~Ybert and L.~Bocquet,
  \emph{Physical Review Letters}, 2012, \textbf{108}, 268303\relax
\mciteBstWouldAddEndPuncttrue
\mciteSetBstMidEndSepPunct{\mcitedefaultmidpunct}
{\mcitedefaultendpunct}{\mcitedefaultseppunct}\relax
\EndOfBibitem
\bibitem[Marchetti \emph{et~al.}(2013)Marchetti, Joanny, Ramaswamy, Liverpool,
  Prost, Rao, and Simha]{Marchetti2013}
M.~C. Marchetti, J.~F. Joanny, S.~Ramaswamy, T.~B. Liverpool, J.~Prost, M.~Rao
  and R.~A. Simha, \emph{Reviews of Modern Physics}, 2013, \textbf{85},
  1143--1189\relax
\mciteBstWouldAddEndPuncttrue
\mciteSetBstMidEndSepPunct{\mcitedefaultmidpunct}
{\mcitedefaultendpunct}{\mcitedefaultseppunct}\relax
\EndOfBibitem
\bibitem[Tailleur and Cates(2008)]{Tailleur2008}
J.~Tailleur and M.~Cates, \emph{Physical Review Letters}, 2008, \textbf{100},
  218103\relax
\mciteBstWouldAddEndPuncttrue
\mciteSetBstMidEndSepPunct{\mcitedefaultmidpunct}
{\mcitedefaultendpunct}{\mcitedefaultseppunct}\relax
\EndOfBibitem
\bibitem[Jiang \emph{et~al.}(2010)Jiang, Yoshinaga, and Sano]{Jiang2010}
H.~R. Jiang, N.~Yoshinaga and M.~Sano, \emph{Phys. Rev. Lett.}, 2010,
  \textbf{105}, 268302\relax
\mciteBstWouldAddEndPuncttrue
\mciteSetBstMidEndSepPunct{\mcitedefaultmidpunct}
{\mcitedefaultendpunct}{\mcitedefaultseppunct}\relax
\EndOfBibitem
\bibitem[Golestanian(2012)]{Golestanian2012}
R.~Golestanian, \emph{Phys. Rev. Lett.}, 2012, \textbf{108}, 038303\relax
\mciteBstWouldAddEndPuncttrue
\mciteSetBstMidEndSepPunct{\mcitedefaultmidpunct}
{\mcitedefaultendpunct}{\mcitedefaultseppunct}\relax
\EndOfBibitem
\bibitem[Cohen and Golestanian(2014)]{Cohen2014}
J.~A. Cohen and R.~Golestanian, \emph{Physical Review Letters}, 2014,
  \textbf{112}, 068302\relax
\mciteBstWouldAddEndPuncttrue
\mciteSetBstMidEndSepPunct{\mcitedefaultmidpunct}
{\mcitedefaultendpunct}{\mcitedefaultseppunct}\relax
\EndOfBibitem
\bibitem[Soto and Golestanian(2014)]{Soto2014}
R.~Soto and R.~Golestanian, \emph{Physical Review Letters}, 2014, \textbf{112},
  068301\relax
\mciteBstWouldAddEndPuncttrue
\mciteSetBstMidEndSepPunct{\mcitedefaultmidpunct}
{\mcitedefaultendpunct}{\mcitedefaultseppunct}\relax
\EndOfBibitem
\bibitem[Soto and Golestanian(2015)]{Soto2015}
R.~Soto and R.~Golestanian, \emph{Phys. Rev. E}, 2015, \textbf{91},
  052304\relax
\mciteBstWouldAddEndPuncttrue
\mciteSetBstMidEndSepPunct{\mcitedefaultmidpunct}
{\mcitedefaultendpunct}{\mcitedefaultseppunct}\relax
\EndOfBibitem
\bibitem[Saha \emph{et~al.}(2014)Saha, Golestanian, and Ramaswamy]{Saha2014}
S.~Saha, R.~Golestanian and S.~Ramaswamy, \emph{Physical Review E}, 2014,
  \textbf{89}, 062316\relax
\mciteBstWouldAddEndPuncttrue
\mciteSetBstMidEndSepPunct{\mcitedefaultmidpunct}
{\mcitedefaultendpunct}{\mcitedefaultseppunct}\relax
\EndOfBibitem
\bibitem[Banigan and Marko(2016)]{Banigan2016}
E.~J. Banigan and J.~F. Marko, \emph{Phys. Rev. E}, 2016, \textbf{93},
  012611\relax
\mciteBstWouldAddEndPuncttrue
\mciteSetBstMidEndSepPunct{\mcitedefaultmidpunct}
{\mcitedefaultendpunct}{\mcitedefaultseppunct}\relax
\EndOfBibitem
\bibitem[Z{\"{o}}ttl and Stark(2014)]{Zottl2014}
A.~Z{\"{o}}ttl and H.~Stark, \emph{Phys. Rev. Lett.}, 2014, \textbf{112},
  118101\relax
\mciteBstWouldAddEndPuncttrue
\mciteSetBstMidEndSepPunct{\mcitedefaultmidpunct}
{\mcitedefaultendpunct}{\mcitedefaultseppunct}\relax
\EndOfBibitem
\bibitem[Matas-Navarro \emph{et~al.}(2014)Matas-Navarro, Golestanian,
  Liverpool, and Fielding]{Matas-Navarro2014}
R.~Matas-Navarro, R.~Golestanian, T.~B. Liverpool and S.~M. Fielding,
  \emph{Phys. Rev. E}, 2014, \textbf{90}, 032304\relax
\mciteBstWouldAddEndPuncttrue
\mciteSetBstMidEndSepPunct{\mcitedefaultmidpunct}
{\mcitedefaultendpunct}{\mcitedefaultseppunct}\relax
\EndOfBibitem
\bibitem[Dey \emph{et~al.}(2015)Dey, Zhao, Tansi, M{\'{e}}ndez-Ortiz,
  C{\'{o}}rdova-Figueroa, Golestanian, and Sen]{Dey2015}
K.~K. Dey, X.~Zhao, B.~M. Tansi, W.~J. M{\'{e}}ndez-Ortiz, U.~M.
  C{\'{o}}rdova-Figueroa, R.~Golestanian and A.~Sen, \emph{Nano Letters}, 2015,
  \textbf{15}, 8311\relax
\mciteBstWouldAddEndPuncttrue
\mciteSetBstMidEndSepPunct{\mcitedefaultmidpunct}
{\mcitedefaultendpunct}{\mcitedefaultseppunct}\relax
\EndOfBibitem
\end{mcitethebibliography}
\end{document}